\documentstyle[12pt,epsfig]{article}
\begin{document}
\def\qup{q_{\uparrow}}
\def\qdown{q_{\downarrow}}
\def\Gup{G_{\uparrow}}
\def\Gdown{G_{\downarrow}}
\def\half{{\textstyle{1\over 2}}}
\def\GeV{{\rm GeV}}
\def\MeV{{\rm MeV}}
\begin{titlepage}
\vspace*{-1cm}
\begin{flushright}
DTP/94/38   \\
May 1994 \\
\end{flushright}                                
\vskip 1.cm
\begin{center}                                                                  
{\Large\bf
Spin-dependent Parton Distributions from            \\[2mm]
Polarized Structure Function Data}
\vskip 1.cm
{\large T.~Gehrmann}
\vskip .2cm
{\it Department of Physics, University of Durham \\
Durham DH1 3LE, England }\\
\vskip   .4cm
and
\vskip .4cm
{\large  W.J.~Stirling}
\vskip .2cm
{\it Departments of Physics and Mathematical Sciences, University of Durham \\
Durham DH1 3LE, England }\\
\vskip 1cm                                                                    
\end{center}                                                                    
\begin{abstract}
In the past year, polarized deep inelastic scattering
experiments at CERN and SLAC
have obtained structure function measurements off proton, neutron and
deuteron targets at a level of precision never before achieved.
The measurements can be used to test the Bjorken and Ellis-Jaffe sum rules,
and also to obtain information on  the  parton distributions in 
polarized nucleons.
We perform a global leading-order QCD fit to the proton deep inelastic
data in order
to extract the spin-dependent parton distributions. By using
 parametric forms which are consistent with theoretical expectations at large
 and small $x$, we find that the quark distributions are now rather well
 constrained. We assume that there is no significant intrinsic
polarization of the strange quark sea.  The data are then  consistent with a
 modest amount of the proton's spin carried by the gluon, although the shape
 of the gluon distribution is not well constrained, and several qualitatively
 different shapes are suggested. The spin-dependent distributions we obtain
 can be used as input to phenomenological studies for future polarized
 hadron-hadron and lepton-hadron colliders.
\end{abstract}                                                                
\vfill
\end{titlepage}                                                                 
\newpage                                                                        
\section{Introduction}
In the last few years there has been a resurgence of interest in the spin
structure of the nucleon. This was largely initiated by the measurements
of the polarized structure function $g_1^p$ by the SLAC-Yale \cite{bau83}
 and EMC \cite{emc89}
collaborations. In the `naive' parton model, $g_1$ can, like the
unpolarized structure function $F_1$, be expressed in terms of the
probability distributions for finding quarks with spin parallel or
antiparallel to the longitudinally polarized parent proton:
\begin{eqnarray}
F_1(x,Q^2) &=& \half \sum_q\; e_q^2\;  [q(x) +\bar q(x) ]  \\
g_1(x,Q^2) &=& \half \sum_q\; e_q^2\;  [\Delta q(x) +\Delta\bar q(x) ]\; ,
\label{naiveg1}
\end{eqnarray}
where
\begin{equation}
q = \qup + \qdown\; , \quad \Delta q = \qup-\qdown \; .
\end{equation}
The renewed interest in the subject was triggered  by
the first precision measurement of the
integral of $g_1^p$ by the  EMC collaboration \cite{emc89},
\begin{equation}
\Gamma_{\rm 1}^{p}(\mbox{EMC/SLAC}) \equiv \int_0^1 dx\; g_1^p(x,Q^2)
= 0.126 \pm 0.010 \pm 0.015
\quad (Q^2 = 10\ \GeV^2) \; ,
\end{equation}
which was significantly lower than the `Ellis-Jaffe sum rule'
value of 0.18 \cite{ell74a}.
This latter prediction is obtained by assuming that the net contribution
of strange quarks to the proton spin is negligible. In the context of
this model, 
$\eta_u$ and $\eta_d$, the net spin carried by up and down quarks 
respectively,  can be obtained from the $\beta$-decay rates
of the octet hyperons:
\begin{eqnarray}
\eta_u \equiv \int_0^1 dx \Delta u(x) & = & 2F \nonumber \\
\eta_d \equiv \int_0^1 dx \Delta d(x) & = & F - D \; .
\end{eqnarray}
Attributing the difference between the Ellis-Jaffe prediction and the
SLAC-EMC measurement to a non-zero strange sea polarization leads
to the rather unusual result that the net spin carried by the
quarks,
\begin{equation}
\eta_\Sigma \equiv \int_0^1 dx\;  \Delta \Sigma(x)
= \int_0^1 dx\; [\Delta u + \Delta d + \Delta s ]\; ,
\end{equation}
is very small. This  so-called `spin crisis' precipitated an enormous amount
of theoretical discussion  -- a clear and comprehensive review
can be found in Ref.~\cite{rey93}.

One of the most compelling explanations for the violation of the
Ellis-Jaffe sum rule is that a substantial amount of the proton's
spin is carried by {\it gluons}. As first pointed out in \cite{alt88},
the polarized gluon contributes
to $g_1$ via the $\gamma_5$-triangle anomaly.
Thus in perturbative QCD, the naive parton
model result of Eq.~(\ref{naiveg1}) is replaced by
\begin{eqnarray}
g_1(x,Q^2) &=& {1\over 2}\; \sum_q\; e_q^2 \; \int_x^1 {dy\over y}
\;  \left[\Delta q(x/y, Q^2) +\Delta\bar q(x/y,Q^2) \right]\; \nonumber \\
& & \times  \left\{\delta(1-y)
+ {\alpha_s(Q^2) \over 2\pi} \Delta C_q(y) + \ldots \right\} \nonumber \\
& & + \; \frac{1}{9} \int_x^1 \; {dy\over y}\;  \Delta G(x/y,Q^2) \; \left\{
n_f\;   {\alpha_s(Q^2) \over 2\pi}\;\Delta C_G(y) +\ldots \right\}
\label{qcdg1}
\end{eqnarray}
where $\Delta G = \Gup - \Gdown $  is the polarized gluon distribution.
Now according to the Altarelli-Parisi evolution equations \cite{alt77},
\begin{eqnarray}
{d \over d \log Q^2} \; \eta_\Sigma(Q^2) & =&  0 +
{\cal O}(\alpha_s^2) \\
{d \over d \log Q^2} \; \alpha_s(Q^2)\; \eta_G(Q^2)
&  =&  0 + {\cal O}(\alpha_s^2) \; ,
\label{glueevo}
\end{eqnarray}
which implies that the gluon contribution to $g_1$ is formally of the
same order as the quark contribution. 
The splitting of the structure function into two conserved 
quantities associated with quark and gluon contributions to the net spin
is natural in the sense that it allows the former to be identified (up
to finite mass corrections) with  the SU(6) ``constituent" quarks. 
Retaining only the leading terms in Eq.~(\ref{qcdg1}),
assuming three quark flavours, and using a factorization scheme
where  $ \Delta C_G (y) = -\delta(1-y)$ (see Section~2),
gives
\begin{eqnarray}
g_1^p(x,Q^2) & = & \frac{1}{2} \sum_{q=u,d,s} e_q^2 \left[\Delta
q(x,Q^2) + \Delta \bar{q}(x,Q^2)\right] 
                 - \frac{1}{3} {\alpha_s(Q^2) \over 2\pi}\Delta
G(x,Q^2) ,
\label{eq:g1bare}\\ 
\Gamma_1^p & = & \int_0^1 dx\; g_1^p = \frac{2}{9} \eta_u + \frac{1}{18}
                 \eta_d + \frac{1}{18} \eta_s
	       - \frac{1}{3} {\alpha_s(Q^2) \over 2\pi}\eta_G .
\label{eq:gamma}
\end{eqnarray}

It is now straightforward to
calculate the amount of net gluon spin needed to explain the SLAC-EMC
sum-rule data. For example, with $\eta_s = 0$ and  $\eta_u$, $\eta_d$
again obtained from hyperon decays, we find $\eta_G \sim 5$
at $Q^2 \sim 10\ \GeV^2$ \cite{alt89}. Although this might be considered
a surprisingly large number there is no violation of spin conservation,
since there is also a contribution  to the proton's spin from
 the orbital angular momentum of the partons:
 \begin{equation}
 {1\over 2} = {1\over 2}\eta_\Sigma + \eta_G + \langle L_z \rangle\; .
 \end{equation}
According to Eq.~(\ref{glueevo}), the second and third terms on the right-hand
side increase as $\log Q^2$ in such a way that their sum is constant.

In Ref.~\cite{alt89} -- hereafter referred to as AS -- a simple
leading-order QCD model, with no polarized strange sea at $Q^2 =
Q_0^2 = 4\ \GeV^2$, was used to extract polarized quark and gluon
distributions from the SLAC and EMC data on $g_1^p(x,Q^2)$. Although the
integrated parton distributions were reasonably well determined, the
shapes of the distributions, especially those of the $d$-quark and the
gluon, were poorly constrained.  As a result,
counting-rule and Regge arguments
for the large and small $x$ behaviour had to be invoked.
The main purposes of the exercise were (i) to show that a consistent
set of distributions could be derived, and (ii) to present parton
distributions which could be used for polarized lepton-hadron and
hadron-hadron collision phenomenology.

Since the AS analysis was performed, there has been a dramatic increase
in the amount of polarized structure function data available.
Using a polarized $^3$He target, the E142 collaboration at SLAC
has measured the neutron structure function $g_1^n$ \cite{slac93}.
The SMC collaboration at CERN first measured the deuteron structure
function $g_1^d$ \cite{smc93} and more recently has measured $g_1^p$
\cite{smc94b}, improving
on the earlier SLAC-EMC measurements. These new results allow a test
of another important sum rule due to Bjorken \cite{bjo66}
\begin{equation}
\Gamma_{\rm Bj} \equiv
\int_0^1 dx\; (g_1^p(x,Q^2) - g_1^n(x,Q^2)) = {g_A\over g_V}\; \left[
1 - {\alpha_s(Q^2)\over \pi} + \ldots  \right] .
\end{equation}
This sum rule follows from SU(2) isospin invariance
and is a rigorous prediction of QCD. Any disagreement between the
measured and predicted values would invalidate the QCD-improved parton
model approach on which the present study is based.
The most recent experimental measurement -- using all available data
\cite{smc94b} -- gives
\begin{equation}
\Gamma_{\rm Bj}^{exp} = 0.163 \pm 0.017 \qquad (Q^2 = 5\ \GeV^2) ,
\end{equation}
to be  compared to the third-order QCD prediction \cite{ver91}
\begin{equation}
\Gamma_{\rm Bj}^{th} = 0.185 \pm 0.004 \qquad (Q^2 = 5\ \GeV^2) .
\end{equation}
The agreement is acceptable, especially considering that higher-twist
contributions might still give a small contribution at this $Q^2$.
In our parton distribution analysis, therefore, we shall use isospin
symmetry to relate the distributions in the proton to those in the
neutron.

It is interesting that the new  measurement of the integrated proton
structure function \cite{smc94b}, again using all available data,
\begin{equation}
\Gamma_{\rm 1}^{p}(\mbox{SMC/EMC/SLAC}) = 0.142 \pm 0.008 \pm 0.011
\qquad (Q^2 = 10\ \GeV^2) ,
\label{eq:SMC}
\end{equation}
is larger than the old SLAC-EMC result. This means that our polarized
gluon distribution (which accounts for the difference between the
measured value and the Ellis-Jaffe prediction) will now be somewhat smaller
than previous estimates:  in fact we shall show below 
 that\footnote{using also improved $F$ and $D$ values} $\eta_G \sim 2$,
compared to the AS result $\eta_G \sim 5$.

The goal of the present study is to use the latest available data
to update and improve the AS parton distribution analysis.
Following AS, we adopt the point of view that there should be no significant
polarization of the strange quark sea.
Although the sum-rule measurements give an indication of the size
of $\eta_G$, there is a large arbitrariness in the {\it shape}
of the gluon distribution. We will explore several qualitatively different
shapes allowed by the data. The outcome of the analysis will again  be a set of
distributions which will provide benchmarks for future deep inelastic
scattering experiments and which will be useful for phenomenological
analyses of other types of polarized scattering experiment.

As we have already indicated, the first moments of our parton
distributions are constrained by hyperon decay and sum-rule data.
In the remainder of this section, we describe the input data which we
use, and the corresponding first moments which obtain.

The structure function
$g_1^p$ occurs in the antisymmetric part of the hadronic tensor in deep
inelastic scattering, which
can be expressed as the matrix element
 of the axial vector current between two proton states~\cite{bjo66}.
These matrix elements can be further decomposed  into the SU(3)$_f$
singlet and octet pieces. In order to
compare the sum rules with  experiment and to perform a
$Q^2$ evolution of the polarized parton
densities, we rewrite Eq.~(\ref{eq:g1bare}) as
\begin{equation}
g_1^p  =  \frac{1}{12}\Delta q_3(x,Q^2) + \frac{1}{36}\Delta
q_8(x,Q^2) + \frac{1}{9}\Delta\Sigma(x,Q^2) -
	  \frac{1}{3}\frac{\alpha_s(Q^2)}{2\pi}\Delta G(x,Q^2)
\label{eq:g1p}
\end{equation}
with similar expressions for the neutron and deuteron:
\begin{eqnarray}
g_1^n & = & -\frac{1}{12}\Delta q_3(x,Q^2) + \frac{1}{36}\Delta
q_8(x,Q^2) + \frac{1}{9}\Delta\Sigma(x,Q^2) - 
	    \frac{1}{3}\frac{\alpha_s(Q^2)}{2\pi}\Delta G(x,Q^2)
\label{eq:g1n}\\
g_1^d & = & \frac{1}{36}\Delta q_8(x,Q^2) +
\frac{1}{9}\Delta\Sigma(x,Q^2) -
\frac{1}{3}\frac{\alpha_s(Q^2)}{2\pi}\Delta G(x,Q^2) \; .
\label{eq:g1d}
\end{eqnarray}
Here we have defined  octet and singlet quark distributions:
\begin{eqnarray}
\Delta q_3 & = & \Delta u - \Delta d \nonumber\\
\Delta q_8 & = & \Delta u + \Delta d - 2\Delta s \nonumber\\
\Delta\Sigma & = & \Delta u + \Delta d + \Delta s   \; .
\label{eq:deltas}
\end{eqnarray}
The first moments of (\ref{eq:g1p})-(\ref{eq:g1d}) are:
\begin{eqnarray}
\Gamma_1^p &=& I_3 + I_8 + I_0 -
\frac{1}{3}\frac{\alpha_s(Q^2)}{2\pi}\eta_G \nonumber\\
\Gamma_1^n &=& -I_3 + I_8 + I_0 -
\frac{1}{3}\frac{\alpha_s(Q^2)}{2\pi}\eta_G \nonumber\\
\Gamma_1^d &=& I_8 + I_0 - \frac{1}{3}\frac{\alpha_s(Q^2)}{2\pi}\eta_G .
\label{eq:moments}
\end{eqnarray}
The  axial-vector current matrix elements, $I_3$ and $I_8$,
 are related to the $F$ and $D$ couplings:
\begin{eqnarray}
I_3 & = & \frac{1}{12} (F+D) \nonumber \\
I_8 & = & \frac{1}{36} (3F-D) .
\end{eqnarray}
Assuming $\eta_s=0$  gives $I_0=4I_8$,  leading to separate
predictions for the quark contributions to the  $\Gamma_1$'s~\cite{ell74a}.

Measuring $F$ and $D$ in hyperon $\beta$-decays gives their
values at  a momentum transfer scale of $Q^2 \approx
{\cal O}(0.5\ \GeV^2)$.
However $F$ and $D$, being related to the
octet axial vector current matrix elements, are $Q^2$ dependent. 
To fix $\eta_u$ and $\eta_d$ for our fits, we
incorporate these perturbative corrections by replacing
$F$ and $D$ by  $\tilde{F}$ and $\tilde{D}$, which are obtained by
comparing the parton model expressions for
$I_3$ and $I_8+I_0$ with their values corrected to first order
in perturbative QCD. Demanding that $\tilde{F}$ and $\tilde{D}$  reproduce
the corrected values of $I_3$ and $I_8+I_0$, one has, following
\cite{clo93},
\begin{eqnarray}
\tilde{I}_3 = \frac{1}{12}(\tilde{F}+\tilde{D})& = & \frac{1}{12}(F+D)
\left[ 1-\frac{\alpha_s}{\pi}\right] \nonumber\\
\tilde{I}_8 + \tilde{I}_0 = \frac{5}{36}(3\tilde{F}-\tilde{D})& = &
 \frac{1}{36}(3F-D)\left[(1-\frac{\alpha_s}{\pi})
            + 4(1-\frac{\alpha_s}{3\pi})\right]\; . \label{eq:oct}
\end{eqnarray}
This leads to
\begin{eqnarray}
\tilde{F} & = & (1 - \frac{3\alpha_s}{5\pi})F -
\frac{2\alpha_s}{15\pi}D \nonumber \\
\tilde{D} & = & (1 - \frac{13\alpha_s}{15\pi})D -
\frac{2\alpha_s}{5\pi}F\; .
\end{eqnarray}
Taking $n_f=3$ and $\Lambda=177\ \MeV$ (see below),
we calculate  $\tilde{F}$ and $\tilde{D}$ at the starting point of our
perturbative  evolution
$Q_0^2=4\ \GeV^2$ ($\alpha_s(4\ \GeV^2)=0.2879$),
taking the experimental  values for $F$ and $D$ from Ref.~\cite{clo93}:
\begin{eqnarray}
\tilde{F}&=&0.424 \pm 0.008 \nonumber\\
\tilde{D}&=&0.718 \pm 0.007 \; ,
\end{eqnarray}
which fixes the first moments of the polarized $u$- and $d$-quark distributions:
\begin{eqnarray}
\eta_u & = & 0.848 \pm 0.016 \nonumber \\
\eta_d & = & -0.294 \pm 0.011 . \label{eq:eta}
\end{eqnarray}
Finally, comparing the parton model expression for $\Gamma_1^p$ (\ref{eq:gamma})
with the experimental value (\ref{eq:SMC}) gives
\begin{equation}
\eta_G=1.971\pm 0.929 \; .
\end{equation}
The above result for the moments and the sum rules are summarized
 in Table~(\ref{tab:sums}).

\begin{table}[htb]
\begin{center}
\begin{tabular}{|c|c|r|} \hline
\rule[-1.2ex]{0mm}{4ex}& $\eta_u$ & $0.848\pm 0.016$\\ \cline{2-3}
\rule[-1.2ex]{0mm}{4ex}& $\eta_d$ & $-0.294 \pm 0.011$\\ \cline{2-3}
\rule[-1.2ex]{0mm}{4ex}partons & $\eta_s$ & $0.000$\\ \cline{2-3}
\rule[-1.2ex]{0mm}{4ex}& $\eta_{\Sigma}$ & $0.554\pm 0.019$\\ \cline{2-3}
\rule[-1.2ex]{0mm}{4ex}& $\eta_G$ & $1.971\pm 0.929$\\ \hline
\rule[-1.2ex]{0mm}{4ex}& $\Gamma_1^p$ & $0.142\pm 0.015$\\ \cline{2-3}
\rule[-1.2ex]{0mm}{4ex}$g_1$ & $\Gamma_1^n$ & $-0.048\pm 0.015$\\ \cline{2-3}
\rule[-1.2ex]{0mm}{4ex}$(4\ \GeV^2)$ & $\Gamma_1^d$ & $0.047 \pm
0.015$\\ \cline{2-3}
\rule[-1.2ex]{0mm}{4ex}& $\Gamma_1^p-\Gamma_1^n$ & $0.190 \pm 0.002$\\ \hline 
\end{tabular}
\caption{First moments of the quark and gluon
 parton distributions and of $g_1$}
\label{tab:sums}
\end{center}
\end{table}

The remainder of the paper is organized as follows. In the next section
we discuss the different possible definitions of 
the polarized gluon distribution. In Section~3 we describe our fit 
to the polarized structure function data,  in Section~4 we 
discuss the implications
of our fits for the polarized sea quark distributions at higher
$Q^2$, and in Section~5 we summarize
our results and present our conclusions.

\section{Definition of the polarized gluon distribution}
\label{sec:gluon}
The form of $\Delta C_G(x)$ in Eq.~(\ref{qcdg1}) has been a
matter of some dispute during the past few years \cite{rey93,bod90,vog91}.
Kodaira \cite{kod79} first pointed out  that  the anomalous dimension
of the singlet axial vector current was non-zero.
According to Ref.~\cite{kod79}, the first moment of $\Delta C_G(x)$ is
\begin{equation}
\int_0^1 dx \;\Delta C_G(x)= -1 .
\end{equation}
Based on these ideas, Altarelli and Ross~\cite{alt88} evaluated 
the anomalous gluonic
contribution to $g_1$. A crucial point in this calculation is the
regularization of collinear singularities
  (from $g\to q \bar q$) in the photon-gluon fusion
process. Depending on the regularization procedure used, one  obtains
different expressions for $\Delta C_G(x)$. In Ref.~\cite{alt88} the
physical quark mass was used as a regulator, giving 
\begin{equation}
\Delta C_G(x)=(2x-1)\ln{\frac{1-x}{x}} .
\end{equation}
The scheme dependence of $\Delta C_G(x)$ was further 
discussed by Bodwin and Qiu~\cite{bod90}. They  obtained the surprising
result that, in dimensional
regularization and for finite quark masses, the  first
 moment of $\Delta C_G(x)$  vanishes.  Only the  (unphysical) approach
of assuming a small but finite gluon mass reproduced the result of
\cite{kod79}. A careful reanalysis~\cite{vog91} of the polarized
photon-gluon fusion process showed that in fact this disagreement
 between different
regularization schemes comes from the soft non-perturbative region.
Contributions from this region should be 
absorbed into the light quark density, and should therefore not 
contribute explicitly to $\Delta C_G(x)$. This leaves us with two
different  possible $\Delta C_G(x)$, both with first moment  equal to $-1$.
The first  is obtained by introducing a non-zero gluon mass, while
the second  derives from either  introducing finite quark masses
 or from applying
dimensional regularization, always subtracting the soft contribution:
\begin{eqnarray}
\Delta C_G(x)\ \mbox{($k^2 < 0$)} &=& (2x-1)\left(
\ln{\frac{1}{x^2}}-2\right) , 
\label{cg1} \\
\Delta C_G(x)\ \mbox{(DR, $m^2 \neq 0$)} &=&
(2x-1)\left(\ln{\frac{1-x}{x}-1}\right) .
\label{cg2}
\end{eqnarray}
\begin{figure}
\begin{center}
~\epsfig{file=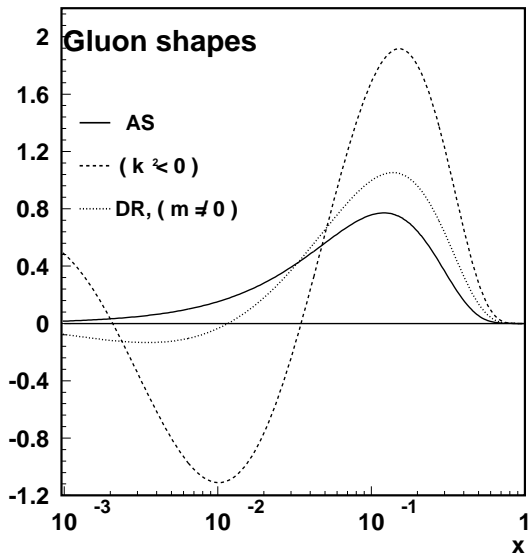,width=7cm}
\caption{$\Delta G$ in different regularization schemes reproducing our
(AS) set A gluon at $Q_0^2$.}
\label{fig:mellin}
\end{center}
\end{figure}
The gluon contribution to $g_1$, Eq.~(\ref{qcdg1}), involves the convolution
of $\Delta C_G(x)$
with $\Delta G$. Rather than tying ourselves to a particular
theoretical scheme, we adopt the AS procedure \cite{alt89}
and define a ``$g_1$-scheme" gluon with $\Delta
C_G(x)=-\delta(1-x)$.  Since only the first moments of the polarized quark and
gluon densities are precisely defined, we are free to fix the higher
moments in this  way to obtain the simplest form for the gluon term in
Eq.~(\ref{qcdg1}). 
Of course, in a consistent next-to-leading order analysis, any change
in the definition of $\Delta C_G$ is compensated by  a corresponding
change in the ${\cal O}(\alpha_s)$
correction to the quark contribution, Eq.~(\ref{qcdg1}),  in such a way
that  the physical quantity $g_1$ is scheme independent. A different
choice of  $\Delta C_G$ will then yield a different gluon {\it and}
a different set of quark distributions.
We are aware of the fact that our  mixture of leading-order quark
and next-to-leading order  gluon contributions in  $g_1$ 
is ambiguous. Nevertheless
we consider the incorporation of the anomalous gluon, although 
formally of
order $\alpha_s$, to be motivated by the scaling behaviour of $\alpha_s
\Delta G$ and by the magnitude of this contribution to $g_1$. 
When  complete 
next-to-leading order corrections become available it will 
of course be necessary
to perform calculations of different processes within a consistent scheme.

 Within our leading-order approach, it is straightforward to
  relate gluon distributions
defined according to different schemes. Suppose that one uses a different
coefficient function, $\Delta\tilde{C}_G(x)$ say, in the
definition of $g_1$. Then
the corresponding polarized gluon distribution $\Delta\tilde{G}(x)$
(i.e. the distribution which reproduces the same $g_1$ at $Q_0^2$) 
is related to ours by 
\begin{equation}
\Delta G(x,Q_0^2) = - \int_x^1 \frac{dy}{y} 
\Delta \tilde{G} \left(\frac{x}{y},Q_0^2 \right)
\Delta \tilde{C}_G(y) .
\label{baregl}
\end{equation} 
This equation can be inverted in Mellin moment space. Applying a
Mellin transformation to Eq.~(\ref{baregl}) and defining 
\begin{eqnarray}
\Delta G(n)&=&\int_0^1 dx\;  x^{n-1} \Delta G(x,Q_0^2) \\
\Delta \tilde{C}_G(n)&=&\int_0^1 dx\;  x^{n-1} \Delta \tilde{C}_G(x) ,
\end{eqnarray}
we obtain
\begin{equation}
\Delta \tilde{G}(x,Q_0^2) = \frac{1}{2\pi i}\int_{a-i\infty}^{ a+i\infty} dn
\;  x^{-n} \;  { \Delta G(n)     \over \Delta \tilde{C}_G(n) } .
\label{invert}
\end{equation}
For reference, we list in Table~\ref{tab:mellin} the Mellin moments
of the coefficient functions  of Eqs.~(\ref{cg1}) and (\ref{cg2}).

In the following section, we will determine  a polarized  gluon
distribution  at $Q_0^2$ from a fit to the structure function data.
The gluon defined in any other scheme can then be obtained by numerical
integration of Eq.~(\ref{invert}). 
As an example,  Figure~\ref{fig:mellin} shows
our gluon set A (see below) together with  the two 
distributions  which  correspond to
the different coefficient functions defined in Eqs.~(\ref{cg1}) and (\ref{cg2}).

\begin{table}[htb]
\begin{center}
\begin{tabular}{|c|c|c|} \hline
\rule[-1.2ex]{0mm}{4ex}$\Delta C_G(x)$ & $\Delta C_G^{(1)}$ & Mellin
transformation $\Delta C_G(n)$\\
\hline\hline 
\rule[-1.2ex]{0mm}{4ex}$-\delta(1-x)$ & $-1$ & $-1$ \\
\rule[-1.2ex]{0mm}{4ex}(AS assumption) & & \\ \hline
\rule[-1.2ex]{0mm}{4ex}$(2x-1)(\ln{\frac{1}{x^2}}-2)$ & $-1$ &
$-2\frac{n^3-n^2+n+1}{n^2(n+1)^2}$ \\
\rule[-1.2ex]{0mm}{4ex}($k^2 < 0$) & &  \\ \hline
\rule[-1.2ex]{0mm}{4ex}$(2x-1)(\ln{\frac{1-x}{x}-1})$ & $-1$ &
$-\frac{n^3+n^2+n+1+(n^3-n)(\gamma_E +
\Psi(n+1))}{n^2(n+1)^2}$\\
\rule[-1.2ex]{0mm}{4ex}(DR, $m^2 \neq 0$) & &   \\ \hline
\end{tabular}
\caption{The different gluon coefficient functions and their Mellin
 transformations}
\label{tab:mellin}
\end{center}
\end{table}

\section{Fit to the polarized structure function data}

\subsection{Parametric form of the fit}
\label{sec:parform}
As a result of over twenty years of experiments, the {\it unpolarized}
parton distributions in the proton  are very precisely known, see for
example Ref.~\cite{mrs93}. Some twenty independent parameters
are needed to specify the various quark and gluon distributions at
$Q_0^2$. In comparison,  measurements of the polarized distributions
are still rather imprecise. It is therefore  necessary to build  in  to the
parametrizations a significant amount of theoretical prejudice.
As measurements improve in the future, these expectations  will be tested and
refinements can be made.

At present, the only constraints on the distributions are (i) the
specification of the first moments of the distributions, as described
in the Introduction and  summarized in Table~\ref{tab:sums}, and (ii)
the requirements of {\it  positivity} of the individual spin-parallel
and spin-antiparallel distributions:
\begin{equation}
f_{\uparrow\downarrow}   = \frac{1}{2}(f \pm \Delta f) > 0 \quad \Rightarrow
\vert \Delta f \vert \leq f\; , \qquad (f=q,G) \; .
\label{eq:pos}
\end{equation}
Therefore,
 the absolute value of a polarized parton disrtibution always has
 to be smaller than
the corresponding unpolarized distribution.
We therefore require a consistent set of leading-order
 unpolarized distributions to provide the bounds -- those of
Ref.~\cite{owe91} are ideal for this purpose. The starting distributions
are, at $Q^2= Q_0^2 = 4\ \GeV^2$,
\begin{eqnarray}
x(u_v + d_v) &=& N_{ud} x^{0.6650} (1-x)^{3.6140} (1 + 0.8673 x) \nonumber\\
x  d_v  &=& N_{d} x^{0.8388} (1-x)^{4.6670} \nonumber\\
x\bar u = x \bar d = x \bar s &=& 0.1515 (1-x)^{7.278} \nonumber \\
x c  & = & 0  \nonumber \\
x G & = & 3.0170 (1-x)^{5.3040}
\label{eq:unpol}
\end{eqnarray}
and the evolution is performed with $\Lambda \equiv \Lambda_{\rm LO}^{(4)}
= 177 \ \MeV$.
The parameters $N_{ud}$ and $N_{d}$ are fixed by the requirement
that $\int_0^1 dx \; u_v = 2 \int_0^1 dx\; d_v = 2$.

For consistency, we choose the same $\Lambda$ and $Q_0$ values as
\cite{owe91}, and similar starting parametrizations. Only our choice
of performing the whole evolution with three quark flavours differs from
\cite{owe91}. 
As explained in the Introduction, we start at $Q_0$
with valence-like forms  for the up and down quarks
and with an unpolarized sea:
\begin{eqnarray}
x\Delta u_{v} & = & \eta_u A_u x^{a_u}(1-x)^{b_u}(1+\gamma_u x) \nonumber\\
x\Delta d_{v} & = & \eta_d A_d x^{a_d}(1-x)^{b_d}(1+\gamma_d x) \nonumber\\
x\Delta\bar u = x \Delta\bar d = x \Delta\bar s  &=& 0      \nonumber \\
x \Delta c  & = & 0  \nonumber \\
x\Delta G & = & \eta_G A_G x^{a_G}(1-x)^{b_G}(1+\gamma_G x) 
\label{eq:form}
\end{eqnarray}
with normalization factors  $A_f$  ($f=q,G$) to ensure that
 $\int_0^1 dx \; \Delta f(x,Q_0^2)  = \eta_f$:
\begin{equation}
A_f^{-1} = \left(1 + \gamma_f \frac{a_f}{a_f+b_f+1}\right)
\frac{\Gamma (a_f) \Gamma (b_f+1)}{\Gamma (a_f+b_f+1)} .
\end{equation}
Non-zero polarized  SU(3) symmetric sea quark distributions 
are generated dynamically for $Q > Q_0$ (see below).
Note that (i)  the positivity constraints (\ref{eq:pos}) demand that the
$b_f$ parameter values are at least as big as the corresponding
unpolarized parameters of Eq.~(\ref{eq:unpol}), and (ii)
the convergence of the first moment integrals requires $a_f > 0$.

Even with these constraints, there are still too many free parameters
in (\ref{eq:form}) for the amount of data available. We therefore
impose additional theoretical constraints at large and small $x$,
based on dimensional counting and coherence/Regge  arguments
 respectively \cite{ellis88,close88,bro90,chi91,bro94}.
A good discussion of these can be found in Ref.~\cite{bro94}.
In the present context, they  can be summarized as   follows:
\begin{enumerate}
\item[{(i)}] The assumption that the small $x$ behaviour
of the quark distributions is controlled by Regge
behaviour  requires $a_u=a_d$.
\item[{(ii)}] Coherence arguments at small $x$ give $a_G^{pol}=a_G^{unpol}+1$
for the gluon distribution.
\item[{(iii)}] Spectator quark counting rules for the $x\to 1$
behaviour suggest $b_d=b_u+1$
and require $b_u$ and $b_d$ to be the same as in the unpolarized case.
\end{enumerate}
We have already mentioned that the most accurate data at present
are for $g_1^p$. This means that the detailed shape of the
$d$-quark and gluon distribution cannot yet be determined.
Thus for the former we impose
$\gamma_u=\gamma_d$, and for the latter we suggest three qualitatively
different forms, defined by
\begin{eqnarray}
\mbox{Gluon A: }\quad \gamma_G & = & 0.0 \nonumber\\
\mbox{Gluon B: }\quad \gamma_G & = & 18.0 \nonumber\\
\mbox{Gluon C: }\quad \gamma_G & = & -3.5
\end{eqnarray}
The reasoning behind these  choices is as follows: $\gamma_G = 0$
(the `standard' form) will give a fit with $\Gup \sim \Gdown$ at large
$x$, which corresponds to a fast-moving gluon being equally likely to have
its spin parallel or antiparallel to the parent proton; $\gamma_G$ large
and positive will lead to $\Gup \gg \Gdown$ at large $x$, i.e.
a fast-moving gluon which is  preferentially polarized in the direction
of the proton's spin; $\gamma_G$ large and negative will give
a large-$x$ gluon whose spin is anticorrelated with that of the proton.
For the latter two cases, we limit the absolute size of the parameter
$\gamma_G$
to ensure positivity of both $\Gup$ and $\Gdown$, and
in each case we fit for the
remaining parameter $b_g$.

Table~\ref{tab:fixed} lists the parameters which are fixed
by the above theoretical considerations. In Section~\ref{sec:fitproc}
 we describe how the remaining parameters
$a_u$, $\gamma_u$ and $b_G$ are determined from the
data.

\begin{table}[htb]
\begin{center}
\begin{tabular}{|c|r|c|} \hline
\rule[-1.2ex]{0mm}{4ex}Parameter & Value &  Comments \\ \hline \hline
\rule[-1.2ex]{0mm}{4ex}$b_u$ & 3.64 & unpolarized $b_u$\cite{owe91}  \\ \hline
\rule[-1.2ex]{0mm}{4ex}$b_d$ & 4.64 & unpolarized $b_d$\cite{owe91} \\ \hline
\rule[-1.2ex]{0mm}{4ex}$a_G$ & 1.000 & coherence arguments\cite{bro90}
\\ \hline
\rule[-1.2ex]{0mm}{4ex}$\eta_u$ & 0.848 & $\eta_u = 2\tilde{F}$ \\ \hline
\rule[-1.2ex]{0mm}{4ex}$\eta_d$ & $-$0.294 & $\eta_d = \tilde{F} -
\tilde{D}$ \\ \hline
\rule[-1.2ex]{0mm}{4ex}$\eta_G$ & 1.971 & from $\Gamma_1^p$ \\ \hline
\end{tabular}
\caption{Parameters fixed by theory}
\label{tab:fixed}
\end{center}
\end{table}

\subsection{Evolution of the  parton distributions}
\label{sec:evo}
Having fixed the starting distributions at $Q_0^2 = 4\ \GeV^2$, the
distributions at higher $Q^2$ are obtained from the
leading-order evolution equations \cite{alt77}:
\begin{eqnarray}
\frac{d\Delta \Sigma(x,Q^2)}{d \ln Q^2} & = & \frac{\alpha(Q^2)}{2\pi}
\int_x^1\frac{dy}{y}\left[\Delta \Sigma(y,Q^2)
     \Delta P_{qq}\left(\frac{x}{y}\right)+\Delta G(y,Q^2)\Delta
P_{qG} \left(\frac{x}{y}\right)\right] \nonumber \\
\frac{d\Delta G(x,Q^2)}{d \ln Q^2} & = & \frac{\alpha(Q^2)}{2\pi} 
 \int_x^1\frac{dy}{y}\left[\Delta \Sigma(y,Q^2)
     \Delta P_{Gq}\left(\frac{x}{y}\right)+\Delta G(y,Q^2)\Delta
P_{GG} \left(\frac{x}{y}\right)\right] \nonumber \\
\frac{d\Delta q_{val}(x,Q^2)}{d \ln Q^2} & = &
\frac{\alpha(Q^2)}{2\pi} \int_x^1\frac{dy}{y}\left[\Delta q_{val}(y,Q^2)
     \Delta P_{qq}\left(\frac{x}{y}\right)\right] \label{eq:evo}
\end{eqnarray}
where the  splitting functions are
\begin{eqnarray}
\Delta P_{qq} & = & \frac{4}{3} \left[\frac{1+z^2}{(1-z)_{+}}+
 \frac{3}{2}\delta(z-1)\right] \nonumber\\
\Delta P_{qG} & = & \frac{1}{2} \left[z^2-(1-z)^2\right]\nonumber\\
\Delta P_{Gq} & = & \frac{4}{3} \frac{1-(1-z)^2}{z}\nonumber\\
\Delta P_{GG} & = & 3\left[(1+z^4)\left(\frac{1}{z}+ 
\frac{1}{(1+z)_{+}}\right)-\frac{(1-z)^3}{z}+\left(
                    \frac{11}{6}-\frac{n_f}{9}\right)\delta(1-z)\right]
\; .
\end{eqnarray}
With the first moments $\Delta P_{qq}^1$ and $\Delta P_{qG}^1$ both
being equal zero
 it follows that
\begin{equation}
\eta_q =  \int_0^1 dx \Delta q_{val}(x,Q^2) \qquad\mbox{and}\qquad
\eta_{\Sigma} = \int_0^1 dx \Delta\Sigma(x,Q^2)
\end{equation}
are both independent of $Q^2$. 
With $\Delta P_{Gq}^1=2$ and $\Delta P_{GG}^1=\beta_0/2$, one finds,
using the renormalization group equation,
\begin{equation}
\label{eq:gluon}
\frac{d}{d \ln Q^2}\alpha_s(Q^2)\eta_G(Q^2) = O(\alpha_s^2)
\qquad\mbox{where} \qquad
\eta_G(Q^2) \equiv \int_0^1 dx \Delta G(x,Q^2) \; ,
\end{equation}
and so  $\Gamma_1^p$, $\Gamma_1^n$ and $\Gamma_1^d$ defined in 
Eq.~(\ref{eq:moments})
 are independent of $Q^2$ in leading order.

\begin{figure}
\begin{center}
~ \epsfig{file=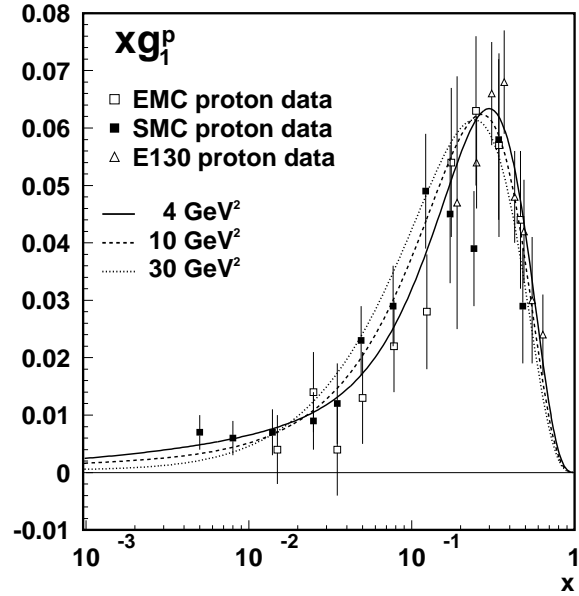,width=7.5cm}
\caption{Fit to the $g_1^p$ structure function with set A gluon}
\label{fig:proton}
\end{center}
\end{figure}

\begin{figure}
\begin{center}
~ \epsfig{file=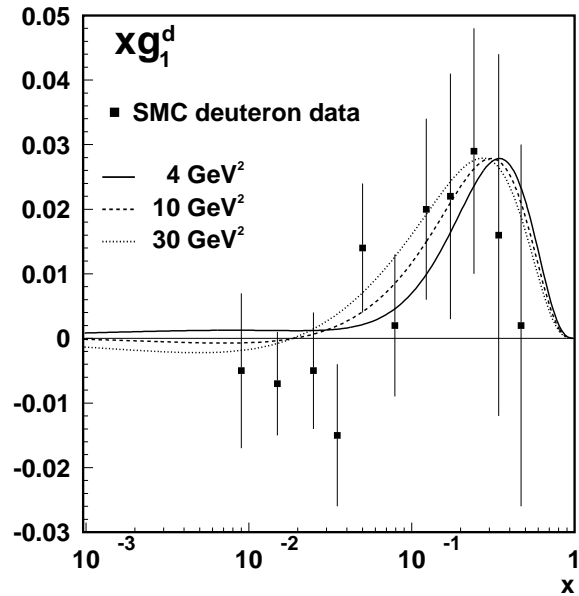,width=7.5cm}
\caption{Prediction for the $g_1^d$ structure function with the set A gluon}
\label{fig:deuteron}
\end{center}
\end{figure}

\subsection{Fitting procedure}
\label{sec:fitproc}
In principle, five sets of $g_1$ structure function data are available
for a global fit: the proton data from SLAC-Yale \cite{bau83}, EMC
\cite{emc89} and SMC \cite{smc94b}, the neutron
data from SLAC-E142 \cite{slac93}, and the deuteron data from SMC
\cite{smc93}.
Our analysis, however, is based on a leading-twist approximation, and it
is therefore necessary to impose a minimum $Q^2$ cut. Based on similar
fits to unpolarized structure function data, see for example \cite{mrs93},
and on recent theoretical estimates \cite{balitsky90,unrau93},
we believe that $Q^2 \geq 4\ \GeV^2$ is sufficient. This means that the
bulk of the SLAC data ($E_{\rm beam} = 30\ \GeV$) falls outside
our fit region.  Only the EMC and SMC  proton data ($E_{\rm beam} \approx
200\ \GeV$) cover a broad range of $x$ for $Q^2 \geq 4\ \GeV^2$.

\begin{figure}
\begin{center}
~ \epsfig{file=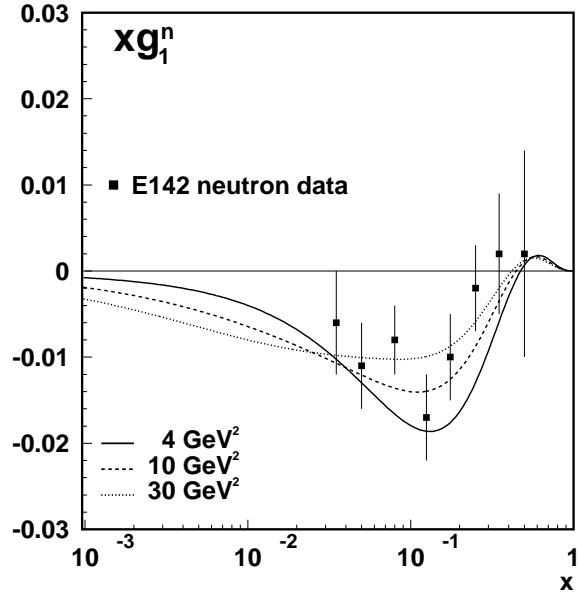,width=7.5cm}
\caption{Prediction for the $g_1^n$ structure function with the set A gluon}
\label{fig:neutron}
\end{center}
\end{figure}

The results of the fit are shown in Figs.~\ref{fig:proton} and
\ref{fig:deuteron}. It turns out that the errorbars on the deuteron data
are so  large that there is no statistically significant constraint.
The fit to $g_1^p$, Fig.~\ref{fig:proton}, is equally good for the
three different types of gluon distribution described in
Section~\ref{sec:parform} (see Table~\ref{tab:fitted}),
 and so only the curves corresponding
to set A are shown. Each data point is fitted at its appropriate
$x$ and $Q^2$ value, but since data from different beam energies
are shown on the same figure, the curves correspond to  fixed $Q^2 = 4,\
10,\ 30\ \GeV^2$. For the SMC and EMC data, these correspond to the
small-, medium- and large-$x$ data respectively. For the SLAC data,
only the largest $x$ data points have $Q^2 \sim 4\ \GeV^2$. The lower
$x$ data points are not used in the fit but are displayed
for completeness. Figure~\ref{fig:neutron} shows our {\it predictions}
for the neutron structure function, together with the
SLAC-E142 data \cite{slac93}. Only the largest $x$ data point
(with $Q^2 = 5.2\ \GeV^2$) can be compared with the theoretical curves.
At lower $x$, we might expect that higher-twist contributions
are important, although it is interesting that the $Q^2 = 4\ \GeV^2$
curve still  gives a fair  description of the data.

\begin{figure}
\begin{center}
~ \epsfig{file=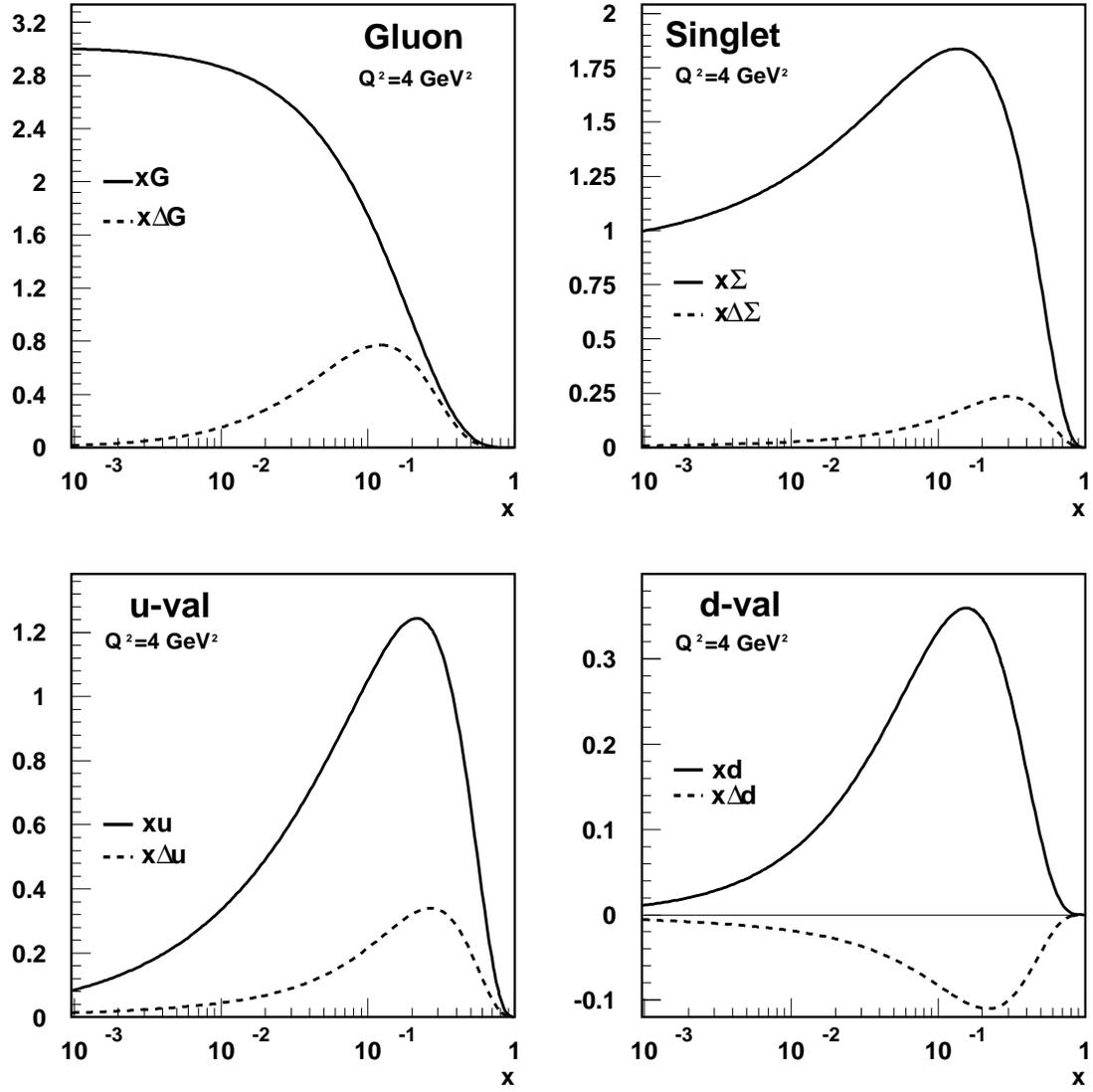,width=16cm}
\caption{The polarized  gluon (set A), quark singlet,
$u_v$ and $d_v$  distributions at $Q_0^2 = 4\ \GeV^2$ obtained
from the fit to the deep inelastic scattering data.}
\label{fig:partons1}
\end{center}
\end{figure}

\begin{figure}
\begin{center}
~ \epsfig{file=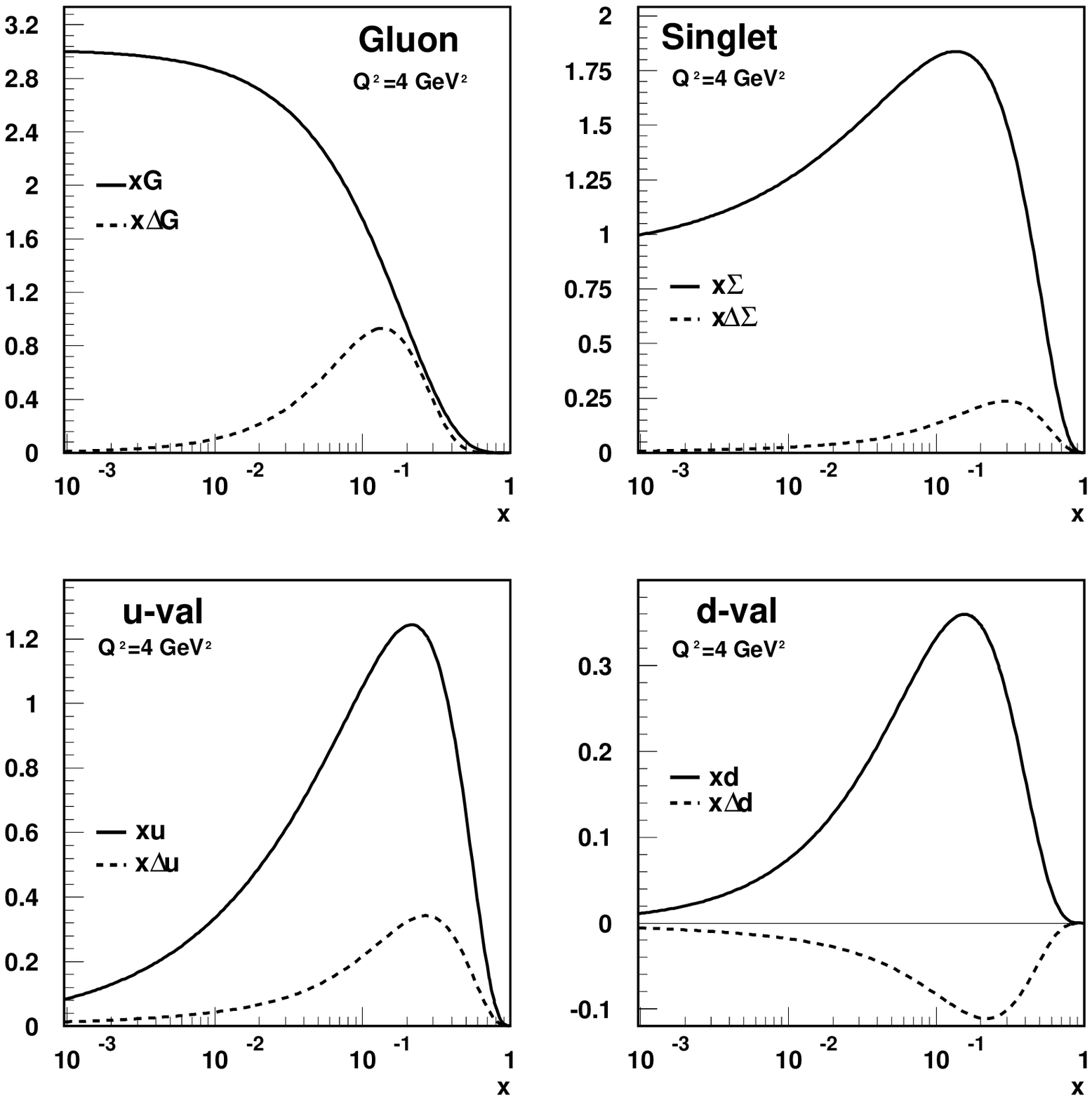,width=16cm}
\caption{As for Fig.~\protect{\ref{fig:partons1}}, but with the set B
gluon distribution.}
\label{fig:partons2}
\end{center}
\end{figure}

\begin{figure}
\begin{center}
~ \epsfig{file=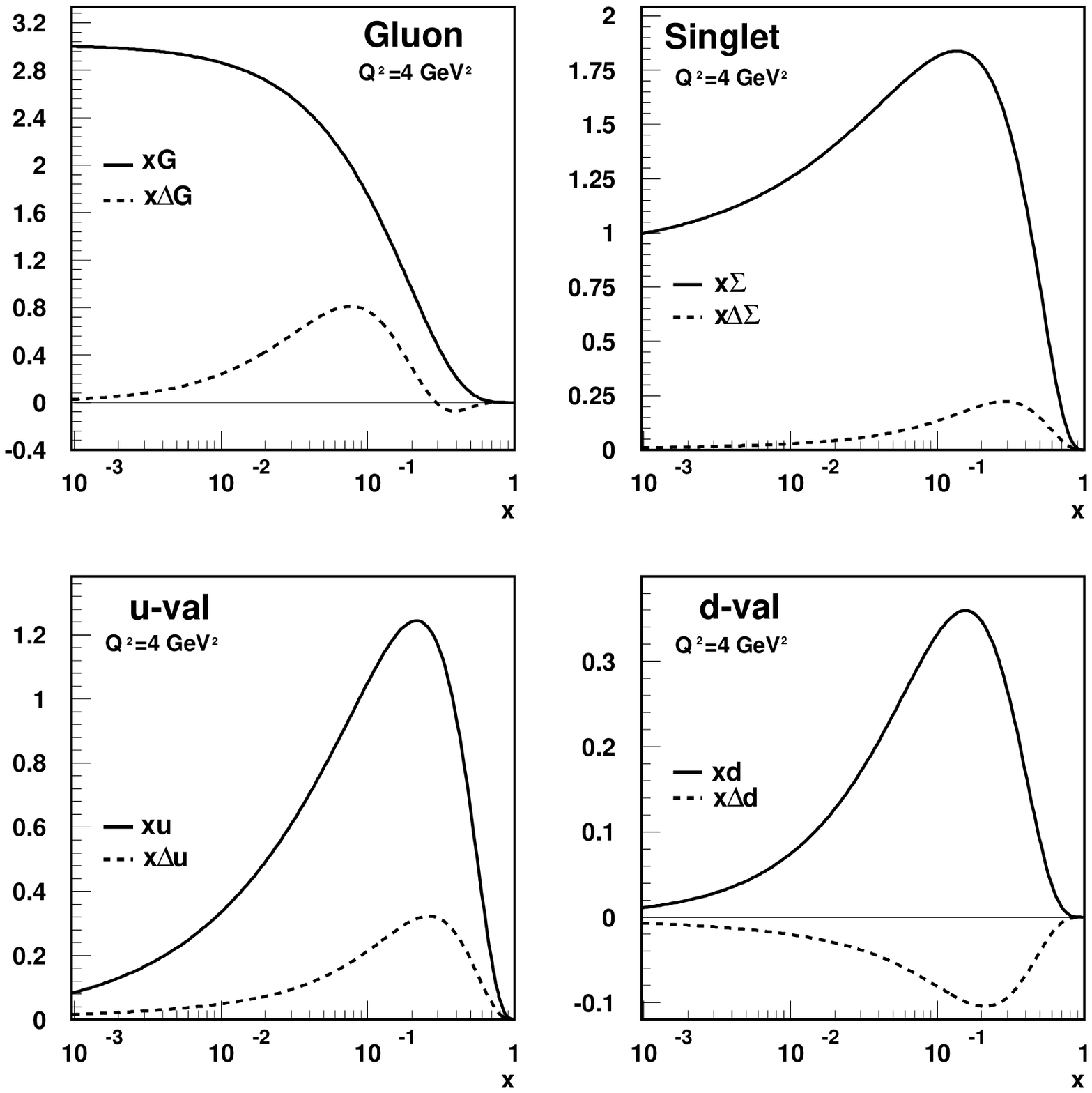,width=16cm}
\caption{As for Fig.~\protect{\ref{fig:partons1}}, but with the set C
gluon distribution.}
\label{fig:partons3}
\end{center}
\end{figure}

The values of the fitted parameters and their associated errors
are listed in Table~\ref{tab:fitted}. Not surprisingly, the quark
distributions are almost independent of the choice of gluon
distribution and the quality of the fit is similar
for sets A, B and C.  Note that the $a_u$  ($=a_d$) parameter
is reasonably constrained to have a value $\simeq 0.45$, close
to the corresponding unpolarized values, see (\ref{eq:unpol}).

\begin{table}[htb]
\begin{center}
\begin{tabular}{|c|r|r|r|c|}\hline
\rule[-1.2ex]{0mm}{4ex}& Gluon A & Gluon B & Gluon C  \\
\rule[-1.2ex]{0mm}{4ex}$\gamma_G$ & 0.0 & 18.0 & $-$3.5  \\ \hline \hline
\rule[-1.2ex]{0mm}{4ex}$\chi^2/{\rm dof}$ & 18.8/30 & 18.9/30 &
20.8/30 \\ \hline
\rule[-1.2ex]{0mm}{4ex}$a_u$ & 0.46$\pm$0.15 & 0.45$\pm$0.12 & 0.43$\pm$0.12
  \\ \hline
\rule[-1.2ex]{0mm}{4ex}$\gamma_u$ & 18.36$\pm$14.49 & 20.22$\pm$12.2 &
17.05$\pm$10.71
 \\ \hline
\rule[-1.2ex]{0mm}{4ex}$b_G$ & 7.44$\pm$3.52 & 11.07$\pm$4.19 & 8.00$\pm$1.59
 \\ \hline
\end{tabular}
\caption{The fitted parameters and $\chi^2/{\rm dof}$  values
corresponding to the three types of gluon distribution. Note that in each
case $a_u = a_d$ and $\gamma_u = \gamma_d$.}
\label{tab:fitted}
\end{center}
\end{table}

Figures \ref{fig:partons1}-\ref{fig:partons3} show the
gluon, quark-singlet, $u$-valence and $d$-valence
distributions at $Q_0^2$ obtained from the fits. For comparison, we show
also the unpolarized distributions of Ref.~\cite{owe91}.

A final comment concerns the shape of the gluon distribution at small
$x$. For all of our sets, the gluonic contribution to $g_1$ is smooth
and negative at small $x$. If one uses a different gluon coefficient
function, for example either of those in Table~\ref{tab:mellin}, then the same
smooth single-signed behaviour can only be reproduced by using an
oscillating gluon, see Fig.~\ref{fig:mellin}. Conversely, a smooth gluon
combined with either of these alternative coefficient functions would
give a gluonic contribution to $g_1$ which changes sign at small $x$.

\section{Predictions for $\Delta \bar{q}$}
\begin{figure}
\begin{center}
~ \epsfig{file=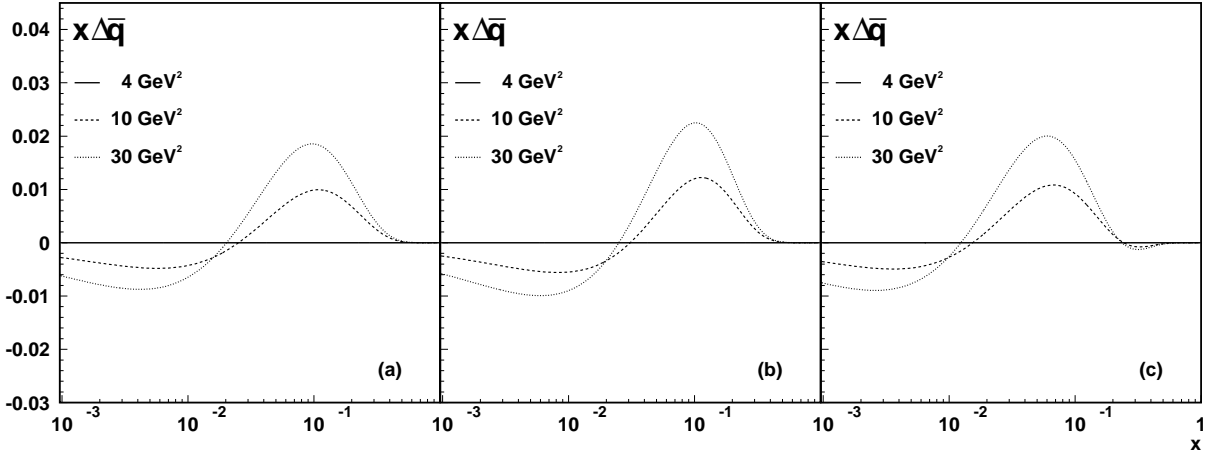,width=16cm}
\caption{The $x\Delta\bar{q}$  distributions at $Q^2 = 4,\ 10,\ 30\
\GeV^2$ corresponding to the  (a) set A, (b) set B, and (c) set C
gluon distributions.}
\label{fig:sea}
\end{center}
\end{figure}

In our model,  a non-zero antiquark distribution
 $\Delta \bar{q}$ is  generated at higher $Q^2$
by  the evolution of the quark singlet, Eq.~(\ref{eq:evo}). The
resulting $\Delta \bar{q}$  is sensitive to
 the initial shape of the gluon. A measurement of the  polarized
sea-quark distribution for $Q^2 > 4\ \GeV^2$ (recall that we generate
an SU(3) symmetric sea)
would therefore probe the gluon distribution. There are several ways
that this could be done. Perhaps the most promising in the short term
is to identify specific flavours of hadrons in the final-state.
The extension of  Eq.~(\ref{naiveg1}) to hadron inclusive structure
functions for $l p \to l H + X$ is
\begin{eqnarray}
F_1^{p\to H}(x,z,Q^2) &=& \half \sum_q\; e_q^2\;
[q(x)D^{q\to H}(z) +\bar q(x)D^{\bar q\to H}(z) ]  \\
g_1^{p\to H}(x,z,Q^2) &=& \half \sum_q\; e_q^2\;
[\Delta q(x)D^{q\to H}(z) +\Delta\bar q(x) D^{\bar q \to H}(z)]\; ,
\label{naivefrag}
\end{eqnarray}
where $D^{q \to H}(z)$ are fragmentation functions. As explained, for
example, in Ref.~\cite{close91}, the identification of $\pi^\pm,\; K^\pm
,\; \ldots$ in the final state allows  the sea-quark polarization
to be probed directly.
Our predictions for the polarized sea-quark distribution are shown
in Fig.~\ref{fig:sea}. Independent of the  choice of gluon, all
the distributions have the common feature of being {\it negative} for small
$x < {\cal O}(0.01)$. At large $x$, the fact that $\Delta G_C$ is less
than zero (see Fig.~\ref{fig:partons3}) is reflected in the
corresponding $\Delta \bar q_C$ distribution.

A preliminary study of the final-state hadron charge asymmetry
in polarized deep inelastic scattering by SMC \cite{wis94} has given
a first indication of the  size  and shape  of the $\Delta \bar{q}$
distribution. Although the  experimental errors are still rather large,
all the distributions of Fig.~\ref{fig:sea} are consistent with these data.

\section{Summary and conclusions}

In this paper we have described an attempt to perform a global fit to
polarized deep inelastic structure function data on proton targets
in order to extract a consistent set of polarized
parton distributions.
We have adopted the point of view that there is no significant
polarization of the sea quarks at low $Q^2$, and therefore that
the gluon carries a substantial fraction of the proton's spin.
We have constrained the net spin carried by the
various parton flavours using deep inelastic sum-rule and hyperon decay
measurements. Recognizing that the quality of the data is not yet
high enough to fully constrain the shapes of the distributions,
we have imposed reasonable theoretical constraints at large and small
$x$. We have performed fits with three qualitatively different gluon
distributions, chosen to reflect the extent to which fast-moving gluons
`remember' the spin of the parent proton.

Our distributions have several applications:
\begin{itemize}
\item[{(i)}] they provide a benchmark for future deep inelastic
structure function measurements --  in particular, measurements
using neutron and deuteron targets,
\item[{(ii)}] they can be compared with theoretical, non-perturbative
model calculations of the spin structure of the proton,
\item[{(iii)}] they can be used to predict cross sections for the
production of various flavours of hadrons in the final state --
$\pi^\pm$ and $K^\pm$ for probing the quark sea, and  the production
of heavy quarks
(in particular $J/\psi$ and open charm)
for probing    the gluon distribution, and
\item[{(iv)}] they can be used as input for the calculation of
hard scattering cross sections in polarized hadron-hadron collisions.
\footnote{More details of these types of processes and
references can be found in Ref.~\cite{rey93}.}
\end{itemize}

In conclusion, polarized lepton-hadron scattering experiments are
beginning to provide precision information on how the spin of the proton
is shared among its parton constituents.
This will be invaluable in testing the predictions of Quantum Chromodynamics
and in designing future experiments. Our distributions should be a
useful tool in both these areas. \footnote{A FORTRAN package
containing the distributions can be obtained by electronic mail from
T.K.Gehrmann@durham.ac.uk}

\section*{Acknowledgements}

We are grateful to R.~Voss and W.~Wislicki for useful
discussions concerning the data. One of us (TG) would like to thank E.~Reya
for drawing his attention to this subject and for numerous clarifying
and motivating discussions. This work was supported in part by
Studienstiftung des deutschen Volkes, Bonn, and by the EEC Programme
``Human Capital and Mobility'', Network ``Physics at High Energy
Colliders'', contract CHRX-CT93-0537 (DG 12 COMA).


\end{document}